\documentclass[10pt]{article}

\usepackage[pdftex]{graphicx}
\graphicspath{{./pdf/}}

\usepackage{csagh}

\usepackage{xcolor}

\newcommand{\linkMS}{\url{https://doi.org/10.5281/zenodo.10992169}}
\newcommand{\linkTOY}{\url{https://doi.org/10.5281/zenodo.10992007}}
\newcommand{\linkProtocol}[1]{Download the \ss{}MACH protocol as PDF to zoom in and explore the protocol (#1).}

\begin{document}
\begin{opening}

\title{Microservices,\newline a Definition Analyzed by \ss{}MACH}

\author[Department of Computer Science,\\
Humboldt-Universität zu Berlin,\\
Berlin, Germany,\\
e-mail: \URL{marcus.hilbrich@informatik.hu-berlin.de}]{Marcus Hilbrich}

\author[Department of Computer Science,\\
Humboldt-Universität zu Berlin,\\
Berlin, Germany,\\
e-mail: \URL{mecquenn@informatik.hu-berlin.de}]{Ninon De Mecquenem}

\begin{abstract}
Managing software artifacts is one of the most essential aspects of computer science. It enables to develop, operate, and maintain software in an engineer-like manner. Therefore, numerous concrete strategies, methods, best practices, and concepts are available. A combination of such methods must be adequate, efficient, applicable, and effective for a concrete project. Eelsewise, the developers, managers, and testers should understand it to avoid chaos. Therefore, we exemplify the \ss{}MACH method that provides software guidance. The method can point out missing management aspects (e.g., the V-model is not usable for software operation), identify problems of knowledge transfer (e.g., how is responsible for requirements), provide an understandable management description (e.g., the developers describe what they do), and some more. The method provides a unified, knowledge-based description strategy applicable to all software management strategies. It provides a method to create a minimal but complete description. In this paper, we apply \ss{}MACH to the microservice concept to explain both and to test the applicability and the advantages of \ss{}MACH.
\end{abstract}




\keywords{\ss{}MACH, microservices, software artefact management, process model, architectural style}

\end{opening}

\section{Introduction}
Managing software artifacts is one of the most essential aspects of computer science.
The question is how to develop, operate, and maintain software~\cite{10.1007/978-3-319-38980-6_25,8267085,10.2307/24020008,Staples2015-STACRA-4,Group1994,TheStandishGroupInternational2015}.
Scientists and industry give different (partial) answers to this question: software process models~\cite{vmodelBund2_3,doi:10.1002/9780470146354.fmatter,rupME,scacchi2001process,scrumG,swManagement,Sommerville2016}, programming paradigms~\cite{10.1145/1113034.1113040,bergs15}, change management~\cite{anderson2010beyond,bernard2011foundations,brewster2012service,cater2005itsmf,edwards1986out,ginger2012embrace,hiatt2006adkar,6612894,marrone2011impact,moen2006evolution_2}, best practices~\cite{10.5555/2616205,Committee2006,ISOIECJTCS2004,Martin08,ISOIECJTCS2018a,ISOIECJTCS2018,JavaI}, and many more~\cite{Bauer1969,its,5587419,elliott2004global,fowler2010domain,4302781,Tracz:1995:DPE:219308.219318,SGM02}.
So, to tackle one challenge, we have various solution strategies, methods, description languages, and suggestions present.
We neither have a uniform solution strategy nor a description language.
For managing concrete software, e.g., in a project~\cite[p.~1]{rupWP}\cite[p.~3,5,14]{scrumG}, software developers, managers, testers, and reviewers are deciding and understanding/learning a management strategy.
Thus, the open question in concrete is:

\textbf{How to review if a software artifact management strategy is suitable ahead of the beginning of the management?}

In principle, a management strategy has to cover at least all relevant aspects of software management.
So, the V-Model is insufficient if a project requires maintaining and operating software.
A software management strategy should not cover additionals, to be minimal.
So, a change management process is insufficient if the decision to create software is already done.
Knowledge transfer is an additional challenge.
If a Kanban-based project decides on an explicit requirement engineering, the results of the requirement engineering are included in the development process.
Therefore, the requirement engineer can deliver a programmer understandable description or participate in the development process.
Requirements are not present in a Scrum process.
 Such a process requires user stories.
 They also describe the same functional properties of software artifacts.

The examples describe the overall challenges.
The different management strategies cover varied aspects, but a project requires managing exactly the project's aspects.
All involved persons have to understand their role in the overall process.
Creating and transferring (by persons, documents, and software artifacts) has to be explicit to enable management.
The different software management strategies use different terms and languages, which makes them hard to understand and combine.

Our solution is a uniform, complete, minimal, and easy-to-apply software guidance.
It applies to all management strategies and their combinations.
\ss{}MACH handled the challenge that an  a priori a unification of the languages used in computer science is not possible~\cite{10.1109/MS.2012.127,WOHLIN2015229}\cite[p.~319]{Sjoberg2008}.
Nevertheless, it is possible to describe the knowledge.
Examples are the description of ontology's of engineering~\cite{10.2307/24020008,Staples2015-STACRA-4}, the work by Popper~\cite{Popper1972-POPOKA,Popper1977-POPTSA}, or other general categorizations~\cite{Staples2015-STACRA-4}.

The \ss{}MACH method~\cite{mh:ssmach,ssp21} provides a unification of the aspects based on a unification of management strategies, an extraction of relevant management aspects, and a systematization of these aspects.
As a result, the \ss{}MACH method enables a software developer to check and specify a concrete management strategy with minimal effort.
The management strategy becomes easily understandable to programmers, project managers, engineers, testers, etc.
The \ss{}MACH method enforces a description of all relevant aspects of software management and identifies missing ones enforce completeness.
It enables review strategies, fosters improvements, is a starting point for academic discussions, and is the basis for systematic comparisons of management strategies.
Therefore, it can describe all kinds of management strategies in a uniform knowledge-based language and avoids chaotic processes.

In this paper, we exemplify the \ss{}MACH method to explain it, provide its advantages, and test its applicability.
Therefore, we have to decide on a management strategy.
Scrum, V-model, or Kanban are describing such management strategies.
They are easily describable by the \ss{}MACH method, so we decide on a more challenging task.

We decided on microservices.
It is a concept, not a management strategy.
Microservices are an answer to various scalability challenges.
They enable large and complex systems by scaling the number of services and development teams.
Microservices allow self-management and agile processes such as Scrum.
Therefore, the developers apply the microservices concept to all parts of the overall system.
It requires all developers, managers, software engineers, etc., to understand the microservice concept and to follow it, so the \ss{}MACH-based description in this paper is helpful.

This paper focuses on using the \ss{}MACH method and not creating the \ss{}MACH concept.
Therefore, we explain the usage of \ss{}MACH on a toy example and analyze microservices.
We decided on microservices because they are well-established and widely used in large-scale industrial applications~\cite{ms_nf,ms_msms,ms_o}.
Such systems have proven to be scalable to support several million users.
Academics described them at various conferences~\cite{ms19Copei,ms19Fritzsch,ms19Gabbrielli,ms19Hausotter,ms19Lu,ms19Maschio,ms19Stein} and discussed them heavily.
Thus, microservices are an answer to actual scalability challenges and an object of academic research.

The scalability of microservices is not limited to the user load.
Also, the development is scalable.
Microservice systems consist of individual microservices~\cite{ms_science_engineering,Dragoni2017,microserviceArchitecture,Wilde2016,ms19Wolff}.
Every service is developed and managed by one team (but a team can have multiple services).
The idea of microservices is to keep a team small, as described by Levis and Fowler~\cite{lewis}.
Most importantly, the teams and the microservices stay small, even if the overall system can scale.
It scales by adding additional microservices and teams.

The small teams provide a set of advantages like less management overhead.
The team members are more productive in a flat hierarchy, and agile software management processes are easy to learn.
The microservices and teams are independent of each other.
Thus, the overall management overhead is reduced.
Nevertheless, too many requests based on service communication or other aspects can hinder the productivity of a small team~\cite{Martin08}.
It is the reason for demanding service and team isolation.
It is an essential factor of microservices.
Based on the isolation, we call our microservice definition a strong one, as opposed to code size, the number of team members, or the used technology as the basis~\cite{ws_7msap,ibm_ms,ms_ms,ms_saas,ms_paas}.

In a microservice system, even if the teams are small and self-managed~\cite{7169442,10.1145/2962695.2962707,scrumG}, a minimum set of rules has to be set.
The teams should not break the microservice system, e.g., by building interfaces to other microservices.
In addition, all management aspects need a description.
Software management includes creating, improving, deploying, and operating software artifacts~\cite{Sommerville2016}. 
Also essential are the documentation and communication of the teams.
All members need to understand and agree to the process to avoid conflicts.

\ss{}MACH~\cite{mh:ssmach,ssp21} documents and defines a software management process.
It aims to check the management strategy to support all relevant management aspects and avoid unnecessary management.
\ss{}MACH is minimalistic, based on scientific groundwork and an ontology of key management aspects.
It provides engineer-like systematics.
In this paper, we give a \ss{}MACH protocol to accomplish all significant aspects of software management in a microservice team.
It describes how to create such a protocol.
So you can adapt it to your concrete process.
As a result, based on the \ss{}MACH protocol, we can demonstrate how the independence of microservices is a solution to many management aspects.

We organize the paper as follows:
An introduction to \ss{}MACH in Sec.~\ref{cap:ssmach}, to give the fundamentals of the scientific method.
In Sec.~\ref{cap:description}, the microservice system is defined and explained.
Sec.~\ref{cap:usecase} describes a use case.
Then (Sec.~\ref{cap:filling}), we explain how to fill the \ss{}MACH protocol (perform the \ss{}MACH method).
With the \ss{}MACH protocol, we can provide observations on the microservice-based process to manage, and we will analyze microservices (Sec.~\ref{cap:advantage}).
We close with a conclusion (Sec.~\ref{cap:conclusion}).

\section{A Short Explanation of \ss{}MACH}
\label{cap:ssmach}
The \ss{}MACH\footnote{Systematic Software Management Approaches
Characterization Helper; \ss{} is the German Eszett. You can read and pronounce it as ``ss''.} method is an approach to define and plan a software-management process, e.g., a software development project.
Therefore, \ss{}MACH defines the management process, gives additional context (meta-information), and describes how to cover essential aspects of software management. (As groundwork, see, e.g.,~\cite{10.1145/1113034.1113040,basili1992software,bernard2011foundations,Stephen,dennis2009systems,Duell1997,DBLP:journals/corr/Fernandez016,fowler2003patterns,4302781,soa1,doi:10.1080/00139157.1996.9930973,omgFP,swManagement,Sommerville2016,7322153}.)
The key aspects are based on an ontology of software engineering and software management strategies and are described based on the vocabulary of knowledge management.

The \ss{}MACH protocol consists of three parts: the definition of management processes, meta-information, and descriptions of the key aspects.
A team should fill in a protocol for each separate management process.
The meta-information defines the team, the filler of the protocol, and additional parts.
The definition of \emph{Our Team} and \emph{Cooperating Teams} are essential for understanding the protocol.
Our team is the group of persons who directly manage the software.
In our example, the team manages one microservice of the overall microservice system.
Cooperating teams are other teams that are intensively involved.
An example would be a dedicated testing team.
In our case study, there is no such team.

\begin{figure}[htp]
\centering
\includegraphics[width=1\textwidth,page=1,trim={31mm 138mm 31mm 36mm},clip,angle=0]{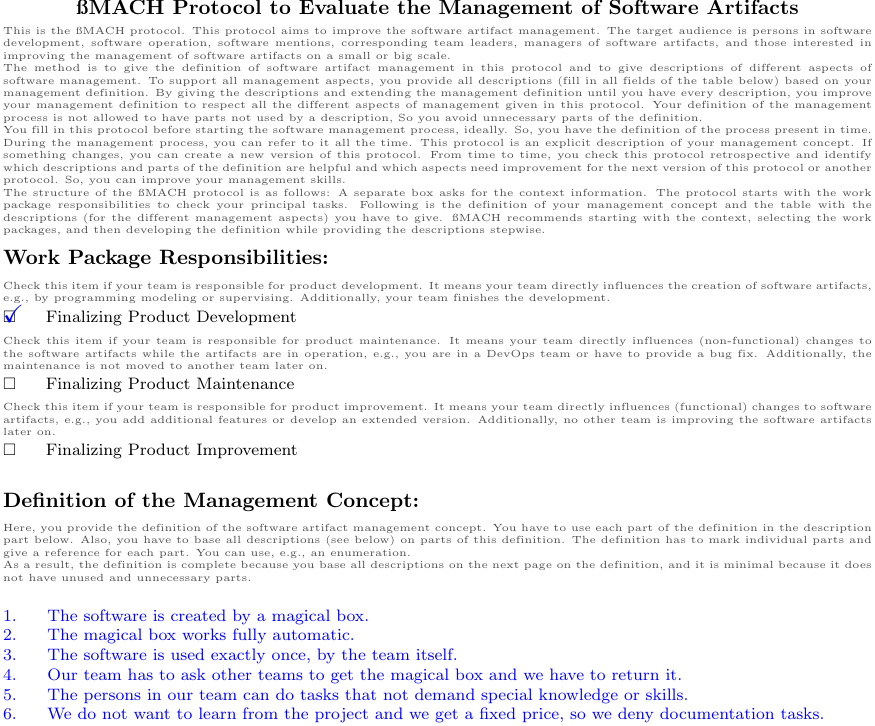}
\caption[]{\label{pic:ssmach:toy:def}Definition of software development toy example in the \ss{}MACH protocol. The definition gives the numbers of the bullet points/parts. 
(See Fig.~\ref{pic:ssmach:toy:keyExp} and \linkTOY{} for more details.)}
\end{figure}

By defining the management process (Fig.~\ref{pic:ssmach:toy:def} and \ref{pic:ssmach:definition}), we provide the guidelines to plan the management.
It can be short and link to additional documents.
For instance, we can reference the Scrum Guide~\cite{scrumG} in the case of Scrum-like management.
The definition should be easy to understand by the target audience.
Usually, this audience is the team. In this case, the readers of the paper.
The definition should be in numbered bullet points.
So it can efficiently describe the key aspects.

\ss{}MACH defines three groups: \emph{our team}, \emph{cooperating teams}, and \emph{externals}.
\emph{Our team} is the group that manages the software, so they have to deliver and operate a microservice (of the microservice system).
Also, \emph{our team} defines and learns the management process in the \ss{}MACH protocol.
\emph{Cooperating teams} are other teams that \emph{our team} can or must cooperate with.
\emph{Our team} can not define how \emph{cooperating teams} work (they do their own management).
\emph{Our team} can change the agreements with them during the management process.
Such teams are, e.g., teams in the same organization.
Because \emph{our team} is working with such teams, \ss{}MACH calls this \emph{Internal}.
The last group is called \emph{External}.
\emph{Our team} can not directly influence such parties, e.g., contract partners.
\emph{Our team} has a defined contract and has to follow it.
Another example is a provider of a library or end-users who use the microservice.

The \ss{}MACH protocol organizes the description of the key aspects of software management in a table.
Fig.~\ref{fig:ssmach:table} provides an example.
The columns define the different aspects of knowledge and information management.
It includes who is doing (column \emph{Roles}), what needs to be known to perform the process (column \emph{Process Knowledge}), and how is the product or aim of the process (column \emph{Product Knowledge}).
This part of \ss{}MACH follows the idea that a product or artifact is created and managed by actors/persons/roles in a process.~\cite{Sommerville2016}.
In addition, \ss{}MACH points out if a piece of knowledge is not present at the beginning (column \emph{Demanded Knowledge}).
Clarifying which knowledge is required is essential as the process needs to find a solution to acquire it during the management.
The last column is called \emph{Process Information}, which defines which information has to be provided by the management process, e.g., working hours for billing, the results of meetings, and delivery protocols.

The rows define the product aspects and the party that influences the aspects.
Previously, we gave the type of parties.
Our team has external parties (marked by the term \emph{Outside}), and our team has cooperating teams (marked by the term \emph{Inside}).
Our team is present by the table and needs to conduct the management process based on the key aspects, given in the table.

The rows in the table represent the \emph{Product Properties}.
The artifacts our team has to develop/manage.
\emph{Interfaces} are the definition of (technical) interfaces of our artifacts to communicate with other systems.
\emph{Dependencies} describe everything our team demands to get from others.
\emph{Responsibilities} give what our team needs to provide to others (e.g., based on contracts or regulations).
Each of the four aspects is present as internal and external.
So, \ss{}MACH defines interfaces to cooperating teams that can be discussed and adapted based on the project needs and fixed interfaces to external parties.
These aspects cannot be influenced directly by our team.

The last row is \emph{External Artifacts}.
It describes that an artifact is taken from another party and included in our project.
It is copied (e.g., use an open-source library).
As a result of copying, it is irrelevant whether it is from a cooperating team or external.
Nevertheless, an external artifact needs management.
Our team needs to know how the artifact works, how we will use it, and other consequences (e.g., based on licensees) the team has to consider.

In addition, \ss{}MACH defines relations of different aspects, abstracted as cells in the table.
One aspect can require another, so knowledge transfer or transformation is required.
A provides relation expresses that an aspect does not need active management.
The aspect is handled/provided as a consequence of managing another one.
For example, forbidding the usage of external artifacts provides a solution to all related management aspects by avoiding any need for management.

\begin{figure}[htp]
\centering
\includegraphics[width=0.94\textwidth,page=2,trim={31mm 161mm 31mm 36mm},clip,angle=0]{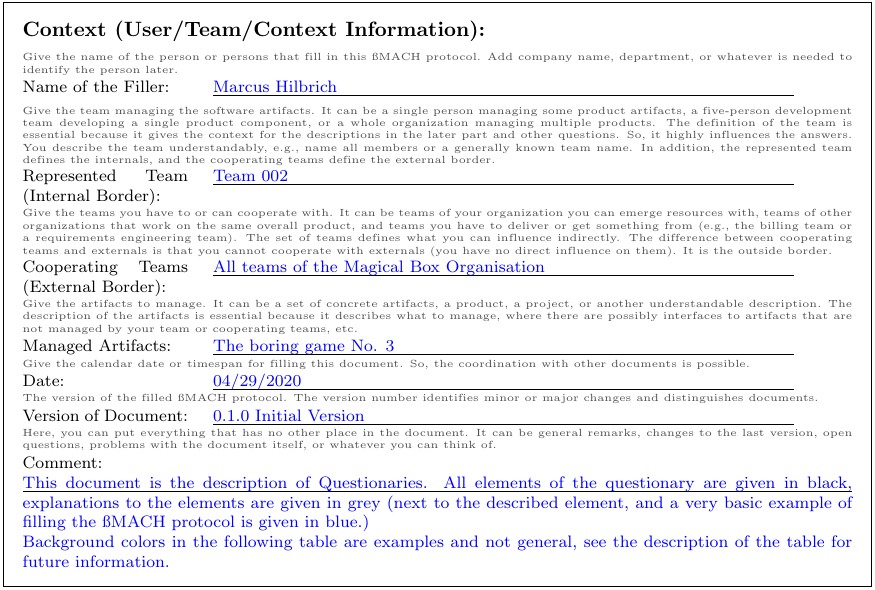}
\caption[]{\label{pic:ssmach:toy:meta}Context or meta-data of software development toy example in the \ss{}MACH protocol. (See Fig.~\ref{pic:ssmach:toy:keyExp} and \linkTOY{} for more details.)}
\end{figure}

\begin{figure}[htp]
\centering
\includegraphics[width=0.94\textwidth,page=2,trim={31mm 127mm 31mm 138mm},clip,angle=0]{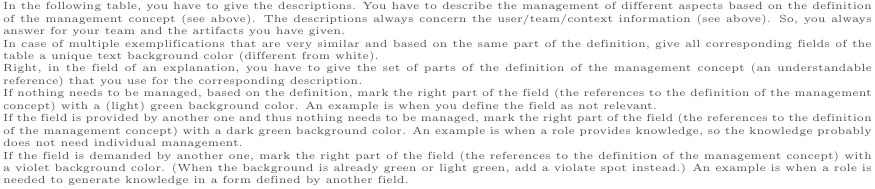}
\caption[]{\label{pic:ssmach:toy:keyExp}The text is part of the hints to the \ss{}MACH protocol. It describes the filling in of the key aspects. The key aspects are separated into Fig.~\ref{pic:ssmach:toy:key1}, \ref{pic:ssmach:toy:key2}, and \ref{pic:ssmach:toy:key3}. \ss{}MACH describes a set of key aspects. Each cell of the table (Fig.~\ref{pic:ssmach:toy:key1}, \ref{pic:ssmach:toy:key2}, and \ref{pic:ssmach:toy:key3}) represents an aspect. The right part of each cell holds the references to the definition in the \ss{}MACH protocol (Fig.~\ref{pic:ssmach:toy:def}). \ss{}MACH defines coloring. Based on the management process, we use light green in the right part of a cell for aspects that do not need active management. Active management means that an aspect is realized without a need for action. The darker green indicates that an aspect is also performed without needing active management but is provided by another. We use violet for aspects, used or required by additional ones. Such an aspect indicates a special interest. Arrows with a peak-end describe a provides-relation. The aspect on the peak is provided by the other. A round end arrow gives a demand relation. The other aspect needs the one at the rounded end. The left part of the call can be colored, too. If the left part of multiple cells uses the same color, the cell's descriptions are equal or very similar. The described aspects in this figure are all based on the other parts of the \ss{}MACH protocol, provided in Fig.~\ref{pic:ssmach:toy:def} and \ref{pic:ssmach:toy:meta}. \linkProtocol{\linkTOY{}}}
\end{figure}

\begin{figure}[htp]
\centering
\includegraphics[width=1\textheight,page=3,trim={43mm 266mm 40mm 51mm},clip,angle=90]{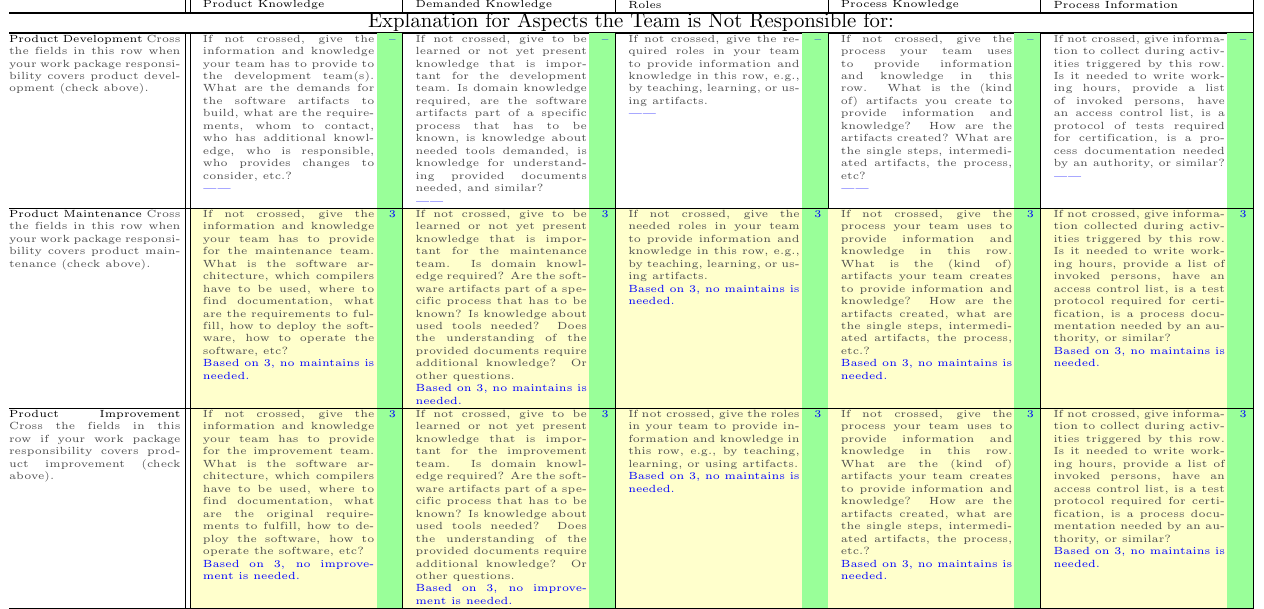}
\caption[]{\label{pic:ssmach:toy:key1}Description of software development toy example based on the \ss{}MACH protocol. Presented is a set of key aspects. (See Fig.~\ref{pic:ssmach:toy:keyExp} and \linkTOY{} for more details.)}
\end{figure}

\begin{figure}[htp]
\centering
\includegraphics[width=1\textheight,page=4,trim={43mm 232mm 40mm 51mm},clip,angle=90]{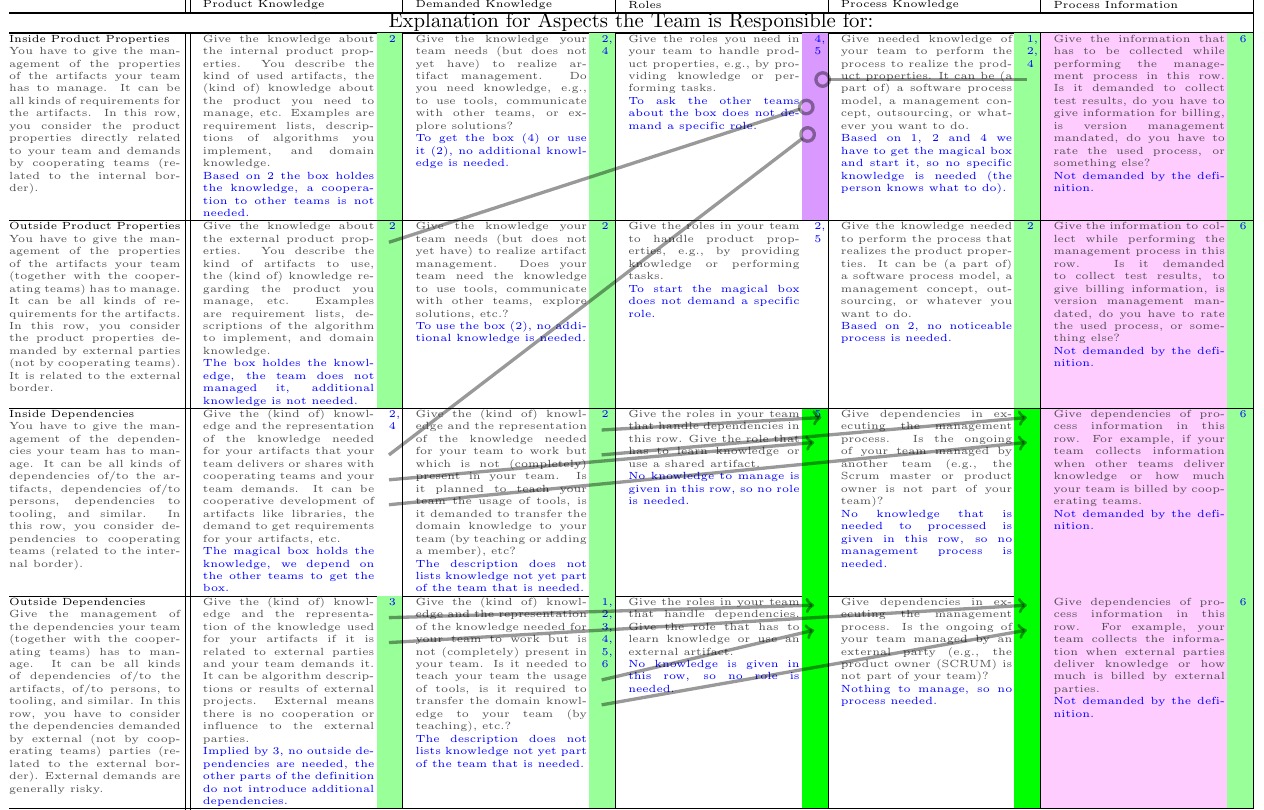}
\caption[]{\label{pic:ssmach:toy:key2}Description of software development toy example based on the \ss{}MACH protocol. Presented is a set of key aspects. (See Fig.~\ref{pic:ssmach:toy:keyExp} and \linkTOY{} for more details.)}
\end{figure}

\begin{figure}[htp]
\centering
\includegraphics[width=1\textheight,page=5,trim={43mm 218mm 40mm 51mm},clip,angle=90]{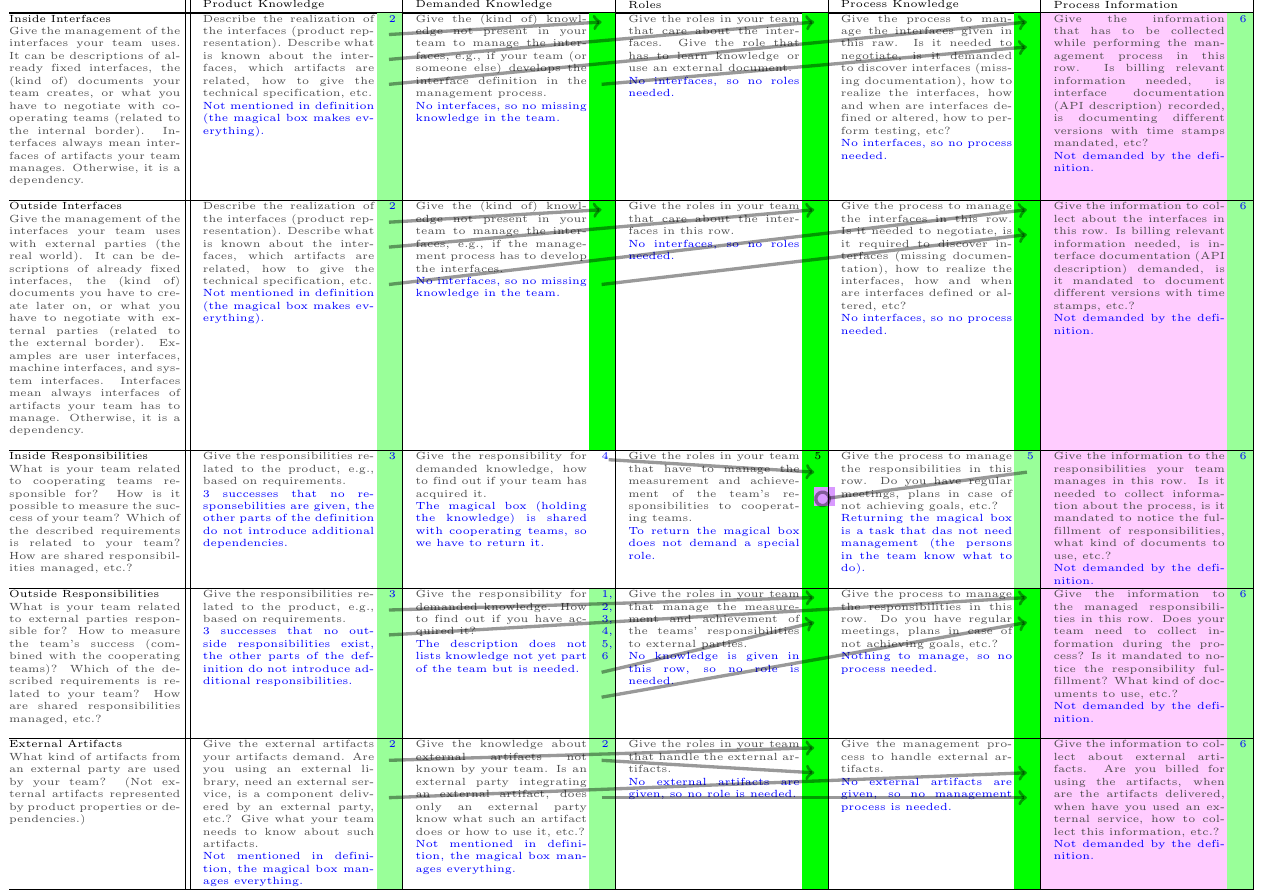}
\caption[]{\label{pic:ssmach:toy:key3}Description of software development toy example based on the \ss{}MACH protocol. Presented is a set of key aspects. (See Fig.~\ref{pic:ssmach:toy:keyExp} and \linkTOY{} for more details.)}
\end{figure}

To exemplify the \ss{}MACH method we provide a toy example in the appendix.
The toy example provides additional explanations for all its parts.
Thereby, it is possible to look up what to fill in the protocol and have a very simple example of a filled protocol.
For this paper, we split the protocol into parts to support printing.
The original protocol consists of an A4 page for the definition of the management plus meta-information and A3 pages for the description of the key aspects.
We give the management definition in Fig~\ref{pic:ssmach:toy:def}, and the meta-information in Fig~\ref{pic:ssmach:toy:meta}.
The table with the key aspects has an initial explanation.
We present it in Fig.~\ref{pic:ssmach:toy:keyExp}.
The original protocol provides one table with all key aspects.
In this paper, we split it into three parts, presented in Fig.~\ref{pic:ssmach:toy:key1}, \ref{pic:ssmach:toy:key2}, and \ref{pic:ssmach:toy:key3}.
For the toy example, we use blue text to visualize everything we (or the team) filled in.
The hints are presented in gray text.
You can download the \ss{}MACH protocol at \linkTOY{} to print it in large or zoom in to read all the details.

The toy example is the following.
A company has a magical box to create software.
So, your challenge is to find out how to get the box.
One team of the company is our team.
This team wants to use the software itself.
Thus, communication, dependencies, management goals, etc., are extremely reduced compared to a realistic project.
The magical box can be interpreted as a simplification of outsourcing.
So, payment circumstances, problems with the outsourcing partner, etc., are removed from the example.

The toy examples use the different relations of key aspects.
The fields in the column process information (Fig. \ref{pic:ssmach:toy:key2} and \ref{pic:ssmach:toy:key3}) are all very similar.
Based on the management definition, no information is recorded.
In each case, the part $6$ of the definition is referenced.
Therefore, the similarity is visualized by the same background color.

The toy example provides different examples of require relations.
One example is from the field inside dependencies / product knowledge to inside product properties / roles.
The dependency describes the need to get the magical box.
This box provides the product properties, and someone (a role in the team) needs to get the box from another team of the company.
As a result, the dependency requires a role to support resolving the dependency.

An example of provide relations is, e.g., present in the row inside dependencies.
The inside dependency is defined by getting the magical box.
That is what you need to know to handle the dependency.
So, it is in the column product knowledge.
This field is resolved by a demand relation described above.
In addition, the magical box does not need extended management.
Because it is so easy, it does not need an additional/extra role to manage it or a process.
Thus, product knowledge provides a solution for field roles and field process knowledge.
In this case, it provides a solution because it can be denied to have a role or a process to manage the inside dependencies.

Later on, we describe the filling in of an \ss{}MACH protocol in detail.
So, we stay with the toy example as a self-explaining, very simplified example.

\section{Strict Definition of Microservices }\label{cap:description}
The paper aims to analyze microservices with the \ss{}MACH method.
Therefore, we defined and described microservices in general, and based on the \ss{}MACH method.
We are not describing a concrete microservice project, and we do not deny that those real-world systems need to find compromises between the strict isolation our definition demands and practical circumstances.
Thus, we do not describe a concrete, but rather a preparation of a management process.
It checks if the microservices concept describes all management aspects defined by \ss{}MACH.
For a real-world management process, we would need a more concrete context and an adaption to give missing descriptions in the microservice concept.

Before using \ss{}MACH, we start with the general description:
The term microservice is not well-defined:
The term is used for SOAs~\cite{ms19Wolff} build of small services~\cite{newman2015building,ms19Tilkov}, for a realization of an organizational structure~\cite{lewis}, as a DevOps concept~\cite{lewis,Wilde2016}, or as architectural style~\cite{ms_science_engineering,Dragoni2017,lewis,microserviceArchitecture,ms19Wolff}.
Our definition focuses on the strict isolation of individual services because isolation can be helpful for management processes~\cite{Stephen,Simon}.

In the following, we provide a clear definition of microservices.
We used definitions stated before (see also~\cite{8498256,microservices19,9860130}), a combination of common definitions and strategies, e.g.,~\cite{ms_science_engineering,Dragoni2017,lewis,microserviceArchitecture,ms19Wolff}.
We present our definition as a pattern:

\noindent\colorbox[gray]{0.9}{\begin{minipage}{0.98\columnwidth}
\paragraph*{Name} Microservices (also called Slice Service Style)  
\paragraph*{Problems to Solve} Solves the need for scalability concerning the system load and the number of persons/teams developing the system.
\paragraph*{Definition} The slice service style is an architectural style where the essential aspects of the system are encapsulated in services (slices, microservices, or vertical services).
These services deliver functionality to end-users and have no (or minimal) dependencies on other slices of the system.
It includes code-sharing, usage of interfaces, sharing of manpower, and management of, e.g., creation, deployment, and operation.
\paragraph*{Consequences} Because of the separation of slices to allow scalability, the software process model needs to be adapted or tailored.
The definition of slices influences the overall system and has to be done globally (e.g., up to the design phase of the waterfall model), while the creation and operation of the slices are independent.
Thus, the (global) software process model has to support independent software development (e.g., by realizing each slice as a DevOps project) and a design or architectural phase at the beginning.
\paragraph*{Drawbacks} Because the independence of slices includes teams and persons, the structure of the organization developing the system needs to be aligned.
In addition, independence reduces the knowledge transfer of the persons of different slices and hinders common reuse techniques.
Especially cross-cutting concerns cannot be managed.
\end{minipage}}

\section{Use Case}\label{cap:usecase}
In the following, we present concrete use cases from the development team's perspective.
We give examples of how the \ss{}MACH protocol can be helpful in concrete and how it is used by the team.
Therefore, we use the microservice example as a basis, but we will also point out differences to a concrete \ss{}MACH protocol.

\subsection{The External Artifact Question}
When our development fills in the \ss{}MACH protocol, they get to the row about external artifacts.
Microservices, as a concept, do not provide a clear and commonly accepted solution strategy.
As a result, our team is pointed to this challenge and needs to make a clear and informed decision.
Typical answers are the following:

\begin{itemize}
    \item To reduce the dependencies on external code, we forbid the usage of external liberties.
    In the \ss{}MACH protocol, we fill in that no knowledge regarding external artifacts exists and no management process is required.
    We make it clear to the team members by adding ``It is forbidden to use external artifacts.'' as an item to the management definition. 
    \item To forbid external libraries creates new challenges.
    Encryption, single sign-on, and logging are forced to be re-implemented.
    This is a high, additional effort and very error-prone.
    Using established, well-tested, and continually supported libraries is a solution strategy.
    In such cases, it demands to know the libraries and check that the licenses are exportable.
    Integrating the liberties requires checking for security issues and update own microservice on demand.
    Therefore, it requires adding a process and role to the \ss{}MACH protocol. 
\end{itemize}

The \ss{}MACH concept forces the team to decide how to handle external artifacts.
The team decides on a strategy and avoids unwanted problems like unmanaged security issues based on outdated libraries.

\subsection{Why Not Use Another Microservice?}
Let us assume we have a running microservice, and our team operates and maintains it.
In this situation, our team gets a new member who proposes to use the other microservices to reduce the code base and increase the functionality.

The team can refer to part $1$ of the definition (Fig.~\ref{pic:ssmach:definition}).
Thus, the new member can understand the current situation.

The \ss{}MACH is not written in stone.
If the situation changes, the protocol can be updated.
In this example, it is discussed to remove part $1$ of the definition.
As a result, all key aspects referring to this part (Fig.~\ref{fig:ssmach:table}) are part of the discussion.
In concrete, cooperation with other teams has to be established and managed.
It is, e.g., needed to have a plan if another service changes the interface or is temporarily unavailable.

Whether it is more effort or risk to manage the relation to other teams/microservices or to not use their services is a decision of our team.
\ss{}MACH demands to describe the plan to foster an informed decision.

\subsection{The Functionality of the Microservice.}
One of the open questions in our example is the functionality of a singular microservice.
In \ss{}MACH, this is mainly a question of product knowledge.
To program and maintain the microservice, our team needs the related product properties (see Fig~\ref{fig:ssmach:table}).
In short, the interfaces the microservice provides to the end-user define the product knowledge (the code base of the microservice).
The code needs to implement the realization of the interfaces.
The responsibilities (the definition of what our microservice has to provide to the end-user) define the interfaces.
A chain of demands relations in the \ss{}MACH protocol (see Fig~\ref{fig:ssmach:table}) represents the knowledge transfer.
The open question in the protocol is who (which role) provides the responsibilities of our service as a system's concern of the overall microservice system.

In our example, we can not answer the system's concerns at all.
The concerns require a concrete system and project.
Without knowing the aims, purpose, or business model of the microservice system, we can not answer.
A real microservice system example has such information available, at least for the overall microservice system.

The \ss{}MACH protocol we provide in this paper is for the development team of one microservice.
Thus, from the viewpoint of this team and the corresponding \ss{}MACH, the knowledge of the partial system's concerns (outside responsibilities) needs to be provided somehow.
If we create a \ss{}MACH protocol for another microservice and another team, we encounter the same problem.
As a result, we demand additional teams that define the business capabilities of the overall microservice system.
In addition, such teams separate the overall system's concerns into individual microservices~\cite{Hasselbring2018}.
To describe such a team in \ss{}MACH is another story.
It requires having an overall business strategy~\cite{Stephen} and dividing~\cite{wolfG} the overall business concerns into individual services.

\section{Filling of the \ss{}MACH Protocol}
\label{cap:filling}
We use the microservice definition to describe the filling of the \ss{}MACH protocol.
The first step is the discussion of the context information.
Then, the definition and the aspects of management are discussed in parallel.
The results are Fig.~\ref{pic:ssmach:definition}, \ref{pic:ssmach:context}, and \ref{fig:ssmach:table}.

\subsection{\ss{}MACH Context}
To fill the \ss{}MACH protocol starts with writing down the context information.
This part of the \ss{}MACH protocol defines other parts.
So, it is a good starting point.
Mostly, the context is very clear and easy to fill.
We know the name and the date a priori.
It is the first version, so we label it as $1.0$.

The \ss{}MACH protocol is filled for a team that manages a microservice, not for the organization that manages the overall microservice system.
We call the team Microservice Team A, A to indicate that other teams of this kind exist.
The team needs more details than a name, so we added an explanation in Fig.~\ref{pic:ssmach:context}.
This also describes the artifacts.
In the pattern definition (Sec.~\ref{cap:description}), the part ``This includes code-sharing, usage of interfaces, sharing of manpower, and management of, e.g., creation, deployment, and operation.'' describes the separation and the artifacts.
The part ``by realizing each slice as a DevOps project'' describes the different teams.

The cooperating teams are mostly independent.
Thus, we could define them as external.
Also, the teams belong to the same organization.
The organization manages the overall microservice project.
It argues against a complete disjoin.
We use the system border of the microservice system as the external border.
The mapping of individual microservices to teams is sufficient to describe it: all teams work on the same microservice system as cooperating teams, even if they are independent.
As a result, the context information of the \ss{}MACH protocol is present in the \ss{}MACH protocol in Fig.~\ref{pic:ssmach:context}.

\subsection{\ss{}MACH Definition and Software Management Aspects}\label{cap:filling:definition}
\ss{}MACH is a method to define and discuss a software management process.
The \ss{}MACH protocol defines the process required by a software development team.
The team is responsible for a microservice.
We start with the work packages.
Afterward, we describe the management process of our team.

\subsubsection{Work Package Responsibilities}\label{cap:definition:responsibil}
Work packages of \ss{}MACH describe if \emph{our team} is responsible for the development, maintenance, and improvement.
The pattern-like definition of microservices (Sec.~\ref{cap:description}) mentions development, maintenance, and improvement.
Development is called ``creation''.
The ``deployment'' is a part of the development and/or maintenance (depending on static deployment or the system changes its deployment).
The ``operation'' is at least part of maintenance and can include improvement.
The mention of DevOps confirms that all work packages are included in the management process.
Thus, the team is responsible for all work packages, and we check them in the \ss{}MACH protocol (Fig~\ref{pic:ssmach:definition}).

\subsubsection{Definition of the Management Process}\label{cap:filling:definition:proceeding}
We have a microservice definition (Sec.~\ref{cap:description}), but it is not a \ss{}MACH protocol.
We need a description where different parts of the definition are easy to reference.
Also, each part should describe one aspect and no mixtures.

To get the definition for \ss{}MACH, the definition from Sec.~\ref{cap:description} is decomposed and recomposed.
We can split the first sentence of the pattern-like description into parts that are candidates for the \ss{}MACH definition:
\begin{itemize}
 \item The naming microservice and the classification as architectural style. 
 \item The representation of system concerns as encapsulated services. 
 \item The services deliver functionality to the end-users. 
 \item Services have no (or minimal) dependencies on each other.
\end{itemize}

The second sentence describes what is included in the services and is independent of other services:
\begin{itemize}
 \item Services have a code base. 
 \item Services have interfaces. 
 \item Services have a team (``manpower''). 
 \item Services persist over development, maintenance, and improvement (``creation, deployment, and operation'').
\end{itemize}

The naming and classification as architectural style do not give the descriptions as needed by the \ss{}MACH protocol.
In addition, we can reorder the items in the description of the system and the microservices:

\begin{enumerate}
\item[I]\label{item:a:ms:filling:system_} The microservice system consists of microservices.
\item[II]\label{item:a:ms:filling:dependence_} Microservices have no (or minimal) dependencies on each other.
\item[III]\label{item:a:ms:filling:concern_} Microservices represent encapsulated system concerns.
\item[IV] Microservices are persisting over development, maintenance, and improvement.
\item[V]\label{item:a:ms:filling:enduser_} Microservices deliver functionality to the end-users.
\item[VI]\label{item:a:ms:filling:interface_} Microservices have interfaces.
\item[VII] Microservices have a code base.
\item[VIII] Microservices have a team.
\end{enumerate}

For the \ss{}MACH protocol, it is only allowed to add needed parts to the management definitions.
Thus, it is a reasonable strategy to develop the definition of the management process by answering the questions about the management aspects in the table of the \ss{}MACH protocol.
This table provides two separate parts.
They are the work the team is not responsible for and the part the team needs to manage directly.
In the following, we give both parts.

\subsubsection{Not Responsible For} 
The team is responsible for product development, maintenance, and improvement.
Thus, we have to cross the fields in the table.
So, we finished the rows of product development, product maintenance, and product improvement.
We do not have to prepare for other teams to overtake the work.
It is typical for DevOps-like strategies.

\subsubsection{Responsible For} 
In the following, we have to provide the descriptions of the software management aspects and fill the table.
By filling the table, we have to refer to the parts of the software management concept.
As the current starting point, this part of the \ss{}MACH protocol is empty.

We start with interfaces. (There are two rows for interfaces in the table.)
We already mentioned interfaces in item VI.
The interfaces are inside interfaces in case cooperating teams use the interfaces.
It would be a kind of dependence that II mostly denies.
Thus, the interfaces are mostly used by externals.
Externals are the end-users according to V.
Item III describes the purpose of the microservice.
Because the pattern-like definition does not mention other communication, it needs to be offered by the interfaces.
So, inside interfaces can be mostly denied.
The product properties of the outside interfaces are a subset of the system concerns.
For the \ss{}MACH definition, we combined items I and II from above to part 1 of the \ss{}MACH definition.
In addition, we combine items III, V, and VI from above as part 2 of the \ss{}MACH definition (Fig~\ref{pic:ssmach:definition}).

We state that internal interfaces are not present.
In short, we deny them.
To deny internal interfaces means we can deny the need for product knowledge.
The team does not need to know anything about nonexisting interfaces.
We denied the other cells in the row, too.
There is no need to explore additional knowledge (demanded knowledge), no management process is needed (process knowledge), no one needs to do something (roles), and the team cannot record information about the nonexisting process (process information).
In other words, based on the fact that no interfaces exist (product knowledge), no role is needed.
In \ss{}MACH, this is a provided relation.
Based on the cell product knowledge, other cells in the row are filled/inferred.
Provide relations are marked by an arrow and according to the coloring of the right side of the cell, as provided by the filled \ss{}MACH protocol in Fig.~\ref{fig:ssmach:table}.

To deny product knowledge based on the independence of services can be directly transferred to the rows inside product properties, inside dependencies, and inside responsibilities.
The arrow for the provided relations and the coloring of the right part of the cells are the same.
We applied the same argumentation to the cell product knowledge in different rows.
\ss{}MACH demands to use the same color, in this case (it is not a relation, so the left part of the cell is colored).
Based on the same argumentation, we used the same color (Fig.~\ref{fig:ssmach:table}).
So, we finished the rows inside interfaces, inside product properties, inside dependencies, and inside responsibilities.

Part 2 (Fig.~\ref{pic:ssmach:definition}) of the \ss{}MACH definition does not only define the product knowledge of outside interfaces.
The interface and the concerns define the outside product properties.
Part 2 defines the product knowledge of the outside responsibilities by a subset of the system concerns, too.
Thus, all three fields get the same color in the \ss{}MACH protocol (left side of the cell).
In addition, we visualize that the three fields depend on each other.
The concern of the system presented by the responsibility is best. The interface is just the technical and organizational presentation of the responsibility.
So, it is dependent on fulfilling the responsibility.
The product property is the realization of the interface.
So, the fields of product knowledge in the rows outside product properties, outside interfaces, and outside responsibilities are defined.

The definition of an architectural style (Sec.~\ref{cap:description}) does not describe how and by whom the artifacts should be managed.
The consequences part of the pattern-like definition is helpful.
It gives the tailoring of the software process model.
We start with the part that describes that DevOps projects are present for each microservice.
It helps to describe additional fields.
The DevOps team has to provide all needed roles, and it is small enough to manage itself.
It gives the roles and the process knowledge for the rows outside product properties and outside interfaces.
Because we gave all descriptions based on the DevOps team, we use the same color for all fields (left part of the cells).
We present the result in Fig.~\ref{fig:ssmach:table}.
Now, we add the DevOps team to the \ss{}MACH description as part 3.

The DevOps team knows how to develop and maintain the product.
Thus, the DevOps team members have product knowledge (row outside product properties).
In other words, the DevOps team builds and maintains the software based on their knowledge/experience (and based on the definition of the interfaces).
Microservices do not give an additional knowledge object.
Therefore, we extended the product knowledge cell in this row and the demand relation (Fig.~\ref{fig:ssmach:table}).

Demanded knowledge (row outside product properties) does not exist.
The \mbox{DevOps} team manages the artifacts.
The outside interfaces present a definition of the product.
It adds two provides relations, so everything is present.

The ``consequence'' section of the pattern-like definition (Sec.~\ref{cap:description}) gives the root of the concerns managed by the team.
The product knowledge in the rows outside responsibilities demands it.
The separation of system concerns is not described (probably given by another team).
Thus, it is demanded knowledge for our team.
How to obtain this knowledge is unclear.
We cannot name the needed process, roles, and process information.
We use question marks and red coloring.
Also, the \ss{}MACH definition is extended by part 4.
The outside interfaces are (mainly) defined by the outside responsibilities. In the case of a concrete end-user, we would need additional aspects, concretization, and adaptions.
The DevOps team handles these interfaces.
These are two provides relations.

The column process information is not directly covered by the pattern-like definition (Sec.~\ref{cap:description}), but the concerns of the system can demand such information (e.g., accounting of used resources to benchmark efficiency).
Thus, our team transfers the (description of) product knowledge to process information for the rows of outside responsibilities, outside interfaces, and outside process properties.

The pattern-like definition does not give or define the row outside dependencies (e.g., to use an external service or external artifacts).
Both can be demanded or forbidden by the system's concerns.
So, a demand relation exists.
Otherwise, we can expect that the DevOps team manages artifacts and dependencies (similar to outside product properties and outside interfaces we give them in the already defined color).
The product information depends on the system concerns (e.g., for the outside interfaces, the system concerns define also the process information directly, but we do not describe it this way).
Our team does not demand additional knowledge.
We expect the DevOps team to have the needed skills and knowledge.
So, we finished the table of \ss{}MACH descriptions (Fig.~\ref{fig:ssmach:table}).

\subsection{Filled \ss{}MACH Protocol}
\label{cap:filling:protocol}\label{cap:observations}
\begin{figure}[htp]
\centering
\includegraphics[width=0.94\textwidth,page=1,trim={26mm 167mm 26mm 46mm},clip]{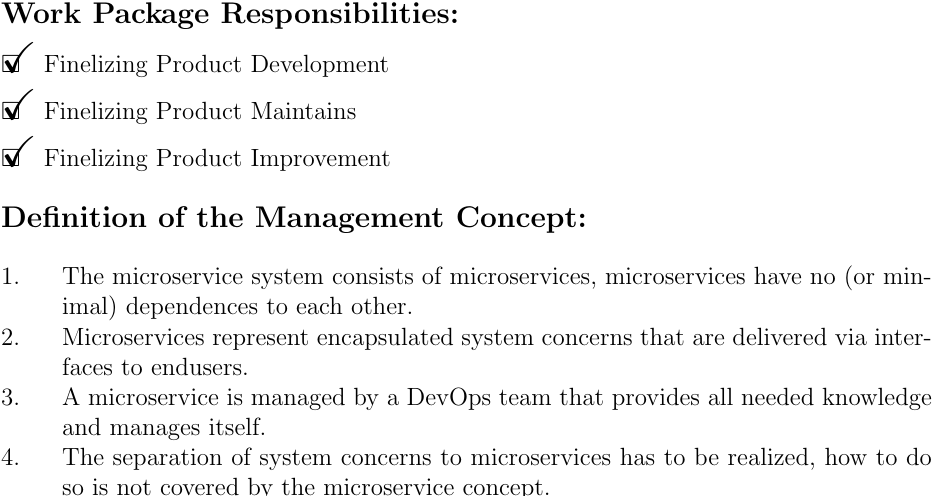}
\caption[]{\label{pic:ssmach:definition}Definition of microservices in the \ss{}MACH protocol. The definition gives the numbers of the bullet points/parts. (See Fig.~\ref{fig:ssmach:table} and \linkMS{} for more details.)}
\end{figure}

We separated the parts of the \ss{}MACH protocol. Fig.~\ref{pic:ssmach:definition} presents the definition, Fig.~\ref{pic:ssmach:context} the meta-data, and Fig.~\ref{fig:ssmach:table} explains the management aspects.

\begin{figure}[htp]
\centering
\includegraphics[width=0.94\textwidth,page=1,trim={26mm 57mm 26mm 142mm},clip]{pdf/ssmachDocumentMS.pdf}
\caption[]{\label{pic:ssmach:context}Context or meta-data of the definition of microservices in the \ss{}MACH protocol. (See Fig.~\ref{fig:ssmach:table} and \linkMS{} for more details.)}
\end{figure}

\begin{figure}[htp]
\centering
\includegraphics[angle=0,width=1\textwidth,page=2,trim={27mm 48mm 44mm 25mm},clip]{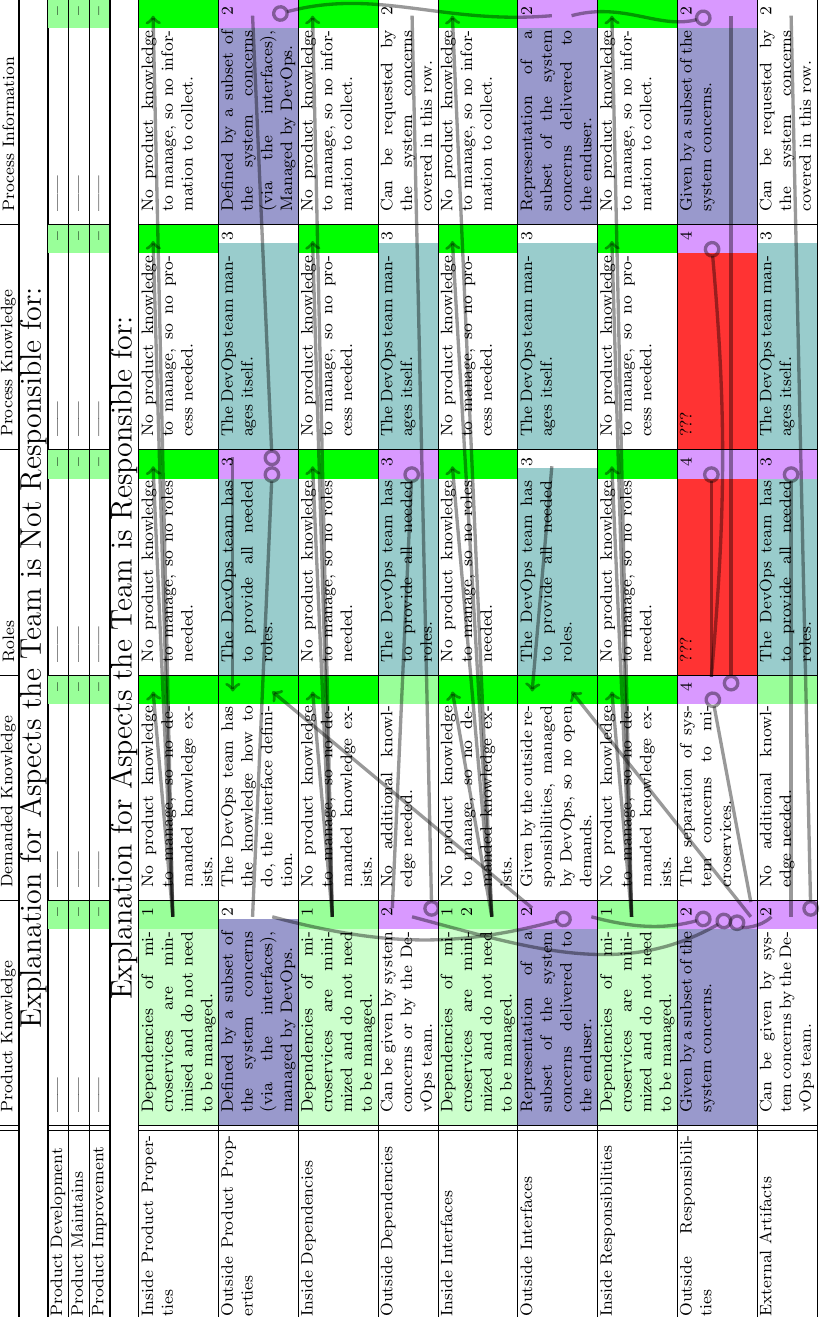}
\end{figure}
\begin{figure}
\caption[]{\label{fig:ssmach:table}Description of microservices based on the \ss{}MACH protocol. \ss{}MACH describes a set of key aspects. Each cell of the table represents an aspect. The right part of each cell holds the references to the definition in the \ss{}MACH protocol (Fig.~\ref{pic:ssmach:definition}). \ss{}MACH defines coloring. Based on the management process, we use light green in the right part of a cell for aspects that do not need active management. Active management means that an aspect is realized without a need for action. The darker green indicates that an aspect is also performed without needing active management but is provided by another. We use violet for aspects used or required by additional ones. Such an aspect indicates a special interest. Arrows with a peak-end describe a provides-relation. The aspect on the peak is provided by the other. A round end arrow gives a demand relation. The other aspect needs the one at the rounded end. The left part of the call can be colored, too. If the left part of multiple cells uses the same color, the cell's descriptions are equal or very similar. The described aspects in this figure are all based on the other parts of the \ss{}MACH protocol, provided in Fig.~\ref{pic:ssmach:definition} and \ref{pic:ssmach:context}. \linkProtocol{\linkMS{}}}
\end{figure}

The \ss{}MACH protocol has only two open, not complete answered cells (Fig.~\ref{fig:ssmach:table}).
The cells describe the separation of concerns of the overall system to isolated microservices.
It is a challenge of microservices~\cite{ws_msnolib,ws_mspfap,msap16,ws_mswq}.

The microservice definition describes many cells, especially product knowledge.
The system concerns are a basis, with many relations in the \ss{}MACH protocol.
The independence of services enables answering inside related rows.

The architectural style does not fully describe roles and processes.
The usage of DevOps answers such questions.

Based on the strict description of microservices (the pattern-like definition, Sec.~\ref{cap:description}), we can fill a \ss{}MACH protocol (Sec.~\ref{cap:filling:protocol}).
Thus, the given definition covers nearly all relevant aspects of software management.
The red-colored cells in the \ss{}MACH protocol point out the open challenge of microservices to define independent concerns of the overall system.

The microservice definition (Sec.~\ref{cap:description}) holds information not present in the \ss{}MACH protocol.
Thus, parts of the definition do not describe software management aspects.

The name of the pattern-like definition is (somehow) represented in the context part of the \ss{}MACH protocol  (Fig.~\ref{pic:ssmach:context}).
The ``problem to solve'' part is not needed to fill the \ss{}MACH protocol.
This part can help to decide whether to use microservices or not.
It is not in the scope of the \ss{}MACH method.
\ss{}MACH helps to understand if all aspects of managing software artifacts are covered.
It is no direct helper to decide to use a specific management method, but it can check different strategies.
So, the aims of the pattern-like definition and \ss{}MACH are different.

The ``definition'' sections of the pattern start with the description in an architectural style.
The \ss{}MACH protocol does not cover it.
So, it describes a pattern property, like the description itself.
The rest of the definition sections cover parts 1 and 2 of the \ss{}MACH definition (Fig.~\ref{pic:ssmach:definition}).

The ``consequences'' part of the pattern-like definition gives two pieces of information.
First, detail of the service separation (part 1 of the \ss{}MACH definition).
Second, the separation of concerns and services has to be realized somehow (part 4 in \ss{}MACH).
Third, individual services are realized by DevOps (part 3 in \ss{}MACH).

The drawback section of the pattern-like definition is not represented by \ss{}MACH.
It describes problems outside our team, and \ss{}MACH does not represent them.
It is an additional different aim of the pattern-like definition and \ss{}MACH.

\section{Results: learning from the \ss{}MACH Protocol}\label{cap:advantage}
We investigate the \ss{}MACH protocol (Sec.~\ref{cap:filling:protocol}):

1) 
We start to look for aspects that do not need active management.
In the \ss{}MACH protocol, such aspects are marked by a green color on the right part of the cell (Fig.~\ref{fig:ssmach:table}).
For the strict definition of microservices, the rows for inside aspects do not need active management.
The reason is also present.
Based on the independence from other services of the same system, no technical (product-based) cooperation with teams of the same microservice system is present.
As a result, the other fields in the rows do not need active management because there is nothing to manage.
There is no need to manage internal relations, a significant advantage.
Fewer communication partners reduce the complexity and the needed team management skills.
The team can concentrate on itself and is probably more productive.
\textbf{\ss{}MACH represents the idea of the strict microservice definition to foster scalability by separating microservices.}

2) 
Based on the provides and demands relations, the \ss{}MACH protocol describes knowledge propagation.
We already mentioned the propagation for the aspects without active management.
The knowledge propagation for active management is interesting for a software engineer.
How is the knowledge transferred and converted, and which knowledge is it?
In Fig.~\ref{fig:ssmach:table}, the starting point is the concerns.
Individual microservices handle the system's business concerns (row outside responsibilities).
The microservice team's responsibilities are the basis for the outside interfaces and the outside dependencies.
Thus, the product properties are indirectly based on the concerns.
In other words, the business concerns of the microservice need to be defined first.
Afterward, the microservice team cares about creating and operating the microservice.
\textbf{The team cares for the microservice.
The \ss{}MACH protocol points out that the team needs a defined business concern as a starting point and then manages the service creation and operation based on the concerns.
Another influence on the team is not present.}
(See Sec.~\ref{cap:usecase} for other management decisions and strategies.)

3) 
Only one kind of description for roles is present in the \ss{}MACH protocol (Fig.~\ref{fig:ssmach:table}).
The roles are not exactly defined.
In other examples of \ss{}MACH protocols, we have seen concrete roles like software developers, architects, and designers.
In Fig.~\ref{fig:ssmach:table}, there is a DevOps team.
The roles this team needs are not fully defined.
Based on the understanding of DevOps, the roles are reasonable to perform the given tasks. 
\textbf{The \ss{}MACH protocol does not point out that the DevOps team needs to adapt to the microservice's business concern.
For a concrete project, we need to define and instantiate the abstract definition of roles.
}

4)
The process knowledge is given by self-management of the DevOps team.
Similar to the roles, this is not concrete.
Nevertheless, DevOps is the idea of small teams and self-management.
A concrete team should give more details on how to perform self-management. 
\textbf{\ss{}MACH points out the DevOps team's self-managed process.
Thus, inadequate influences on the team have to be omitted.
It is also a consequence of the independence of microservices.}

5)
The definition of the management concept in \ss{}MACH (Fig.~\ref{pic:ssmach:definition}) holds four easy-to-read bullet points.
It is very minimal, easy to remember, fast to understand, and interpretation is present and referrable at any time (Fig.~\ref{fig:ssmach:table}). 
Based on our observation, it is very helpful to have an explicit management process.
It makes the process easier, reduces conflicts, and enables improvements.
Also, a change in the process gets obvious, and changes can be explicitly discussed. 
\textbf{The \ss{}MACH protocol is compact, and it is easy to understand the definition of the management process. 
So, the planned process is written down and can be referred to later on.}
(See Sec.~\ref{cap:usecase} for changing the management strategy and updating the protocol.)

6)
The effort to create a \ss{}MACH protocol is not very high and no special knowledge or skills are needed.
To describe microservices, you need to understand microservices.
So, you can create a protocol in about two hours on a whiteboard with the DevOps team.
Afterward, the process is clear to all team members. 
We have also discussed two weeks about a single \ss{}MACH protocol.
We discussed the management process, and we learned a lot.
Based on filling the \ss{}MACH protocol, we identified the gaps in the process, found borderline cases, and nailed down the differences between our idea of the process and the practical doing.
At least based on our observation, the \ss{}MACH method supported us. 
\textbf{A \ss{}MACH protocol is created in some hours and can help to improve the management process.}

7)
\ss{}MACH is a communication helper.
The terms in the protocol support understandability.
The description of the process by one person is easier to understand for the team.
It was even possible to identify misunderstandings between persons.
If two persons answered a management aspect in the \ss{}MACH protocol differently, the process was not yet clear.
\textbf{The definition of terms and the systematics of the \ss{}MACH protocol support the communication of the involved persons and avoid misunderstandings in the management process.%
}

8)
Our definition of microservices is strong for explaining the essentials of the concept.
To allow minimal dependencies is a concession to practical implementations of microservice systems.
Nevertheless, isolation is not easy to realize.
Especially for transforming legacy systems to microservices, the definition gives a goal, not the transformation or an intermediate step.
Thus, a practical realization of a microservice system probably sacrifices strict isolation and decides to manage the consequences instead of dealing with the realization of strict isolation.
\textbf{We use a microservice definition to point out the advantages of strict isolation.
A real-world system uses potbelly less strict definitions with reduced isolation. Especially, the transformation of a monolith into a microservice system will not hold our definition.
In such a case, the \ss{}MACH protocol will look different.}

\section{Conclusion}\label{cap:conclusion}
Based on performing the \ss{}MACH method, we can state two kinds of findings.

First, the \ss{}MACH method is helpful for the analysis of software management processes and supports the management.
The method is systematic and defines terms to describe the process.
So, it supports analyses, such as the understanding of the process by the development team and learning the process by all team members.
Also, the method is easy and fast to perform and thus efficient.

Second, the isolation of individual microservices supports the development team.
The team can avoid many aspects of management.
In addition, the team can perform all the knowledge representation and transformation to develop an individual microservice.
There are no supplementary relations or dependencies to the team.
So, the number of teams can be scaled without overhead to individual teams.

We close this paper by introducing you to fill a \ss{}MACH protocol for your software management process and take value from the method.
To learn more about your management proceeding and how your colleagues understand it.

\section*{Appendix}

\clearpage

\begin{acknowledgements}
Funded by the Deutsche Forschungsgemeinschaft (DFG, German Research Foundation) -- Project-ID 414984028 -- SFB 1404 FONDA
\end{acknowledgements}

\bibliographystyle{cs-agh}
\bibliography{bibliography}

\end{document}